\documentstyle[psfig]{mn}
\newif\ifAMStwofonts
\AMStwofontstrue

\newcommand{\be}{\begin{equation}}
\newcommand{\ee}{\end{equation}}

\def \o0{$\Omega_{0}$}

\ifoldfss
  \ifCUPmtlplainloaded \else
    \NewTextAlphabet{textbfit} {cmbxti10} {}
    \NewTextAlphabet{textbfss} {cmssbx10} {}
    \NewMathAlphabet{mathbfit} {cmbxti10} {} % for math mode
    \NewMathAlphabet{mathbfss} {cmssbx10} {} %  "   "    "
  \fi
  \ifAMStwofonts
    \ifCUPmtlplainloaded \else
      \NewSymbolFont{upmath} {eurm10}
      \NewSymbolFont{AMSa} {msam10}
      \NewMathSymbol{\upi}     {0}{upmath}{19}
      \NewMathSymbol{\umu}     {0}{upmath}{16}
      \NewMathSymbol{\upartial}{0}{upmath}{40}
      \NewMathSymbol{\leqslant}{3}{AMSa}{36}
      \NewMathSymbol{\geqslant}{3}{AMSa}{3E}

    \fi
  \fi
\fi % End of OFSS

\ifnfssone
  \newmathalphabet{\mathit}
  \addtoversion{normal}{\mathit}{cmr}{m}{it}
  \addtoversion{bold}{\mathit}{cmr}{bx}{it}
  \newmathalphabet{\mathbfit} % math mode version of \textbfit{..}
  \addtoversion{normal}{\mathbfit}{cmr}{bx}{it}
  \addtoversion{bold}{\mathbfit}{cmr}{bx}{it}
  \newmathalphabet{\mathbfss} % math mode version of \textbfss{..}
  \addtoversion{normal}{\mathbfss}{cmss}{bx}{n}
  \addtoversion{bold}{\mathbfss}{cmss}{bx}{n}
  \ifAMStwofonts
    \ifCUPmtlplainloaded \else
      \UseAMStwoboldmath
      \makeatletter
      \new@mathgroup\upmath@group
      \define@mathgroup\mv@normal\upmath@group{eur}{m}{n}
      \define@mathgroup\mv@bold\upmath@group{eur}{b}{n}
      \edef\UPM{\hexnumber\upmath@group}
      \new@mathgroup\amsa@group
      \define@mathgroup\mv@normal\amsa@group{msa}{m}{n}
      \define@mathgroup\mv@bold\amsa@group{msa}{m}{n}
      \edef\AMSa{\hexnumber\amsa@group}
      \makeatother
      \mathchardef\upi="0\UPM19
      \mathchardef\umu="0\UPM16
      \mathchardef\upartial="0\UPM40
      \mathchardef\leqslant="3\AMSa36
      \mathchardef\geqslant="3\AMSa3E
    \fi
  \fi
\fi % End of NFSS release 1

\ifnfsstwo
  \DeclareMathAlphabet{\mathbfit}{OT1}{cmr}{bx}{it}
  \SetMathAlphabet\mathbfit{bold}{OT1}{cmr}{bx}{it}
  \DeclareMathAlphabet{\mathbfss}{OT1}{cmss}{bx}{n}
  \SetMathAlphabet\mathbfss{bold}{OT1}{cmss}{bx}{n}
  \ifAMStwofonts
    \ifCUPmtlplainloaded \else
      \DeclareSymbolFont{UPM}{U}{eur}{m}{n}
      \SetSymbolFont{UPM}{bold}{U}{eur}{b}{n}
      \DeclareSymbolFont{AMSa}{U}{msa}{m}{n}
      \DeclareMathSymbol{\upi}{0}{UPM}{"19}
      \DeclareMathSymbol{\umu}{0}{UPM}{"16}
      \DeclareMathSymbol{\upartial}{0}{UPM}{"40}
      \DeclareMathSymbol{\leqslant}{3}{AMSa}{"36}
      \DeclareMathSymbol{\geqslant}{3}{AMSa}{"3E}
    \fi
  \fi
\fi % End of NFSS release 2

\ifCUPmtlplainloaded \else
  \ifAMStwofonts \else % If no AMS fonts
    \def\upi{\pi}
    \def\umu{\mu}
    \def\upartial{\partial}
  \fi
\fi
\begin{document}

\title[Topology of the universe from {\it COBE}-DMR; a wavelet approach]{Topology of the universe from {\it COBE}-DMR; a wavelet approach}
\author[Rocha et al.]{G. Rocha$^1$$^,$$^2$, L. Cay\'on$^3$, R. Bowen$^4$, A. Canavezes$^5$$^,$$^2$, J. Silk$^4$, A. J. Banday$^6$ \and and K. M. G\'{o}rski$^7$$^,$$^8$ \\
$^1$ Astrophysics Group, Cavendish Laboratory,
Madingley Road, Cambridge CB3 0HE, United Kingdom \\
$^2$ Centro de Astrof\'{\i}sica da Universidade do Porto, R. das
Estrelas s/n, 4150-762 Porto, Portugal \\
$^3$Instituto de F\'{\i}sica de Cantabria, Fac. Ciencias, 
Av. los Castros s/n, 39005 Santander, Spain \\
$^4$Department of Physics, Denys Wilkinson Building,
University of Oxford, Keble Road, Oxford OX1 3RH, United Kingdom \\
$^5$ Institute of Astronomy, Madingley Road, Cambridge CB3 0HA, United Kingdom \\
$^6$Max-Planck Institut fuer Astrophysik (MPA), 
Karl-Schwarzschild Str.1, D-85740, Garching, Germany \\
$^7$European Southern Observatory (ESO), 
Karl-Schwarzschild Str.2, D-85740, Garching, Germany \\
$^8$Warsaw University Observatory, Poland \\ 
}

\date{\today}

%\date{Accepted .
%      Received ;
%      in original form }

\pagerange{\pageref{firstpage}--\pageref{lastpage}}
\pubyear{2002}

%\begin{document}

\maketitle

\label{firstpage}

\begin{abstract}

In this paper we pursue a new technique to search for evidence of a finite
Universe, making use of a spherical mexican-hat wavelet decomposition of
the microwave background fluctuations. Using the information provided by
the wavelet coefficients at several scales we test whether compact
orientable flat topologies are consistent with the {\it COBE}-DMR data. We
consider topological sizes ranging from half to twice the horizon size. A
scale-scale correlation test indicates that non-trivial topologies with
appropriate topological sizes are as consistent with the {\it COBE}-DMR data as
an infinite universe.  Among the finite models the data seems to prefer a
Universe which is about the size of the horizon for all but the hypertorus
and the triple-twist torus. For the latter the wavelet technique does not
seem a good discriminator of scales for the range of topological sizes
considered here, while a hypertorus has a preferred size which is $80\%$ of
the horizon. This analysis allows us to find a best fit topological size
for each model, although cosmic variance might limit our ability to
distinguish some of the topologies.

\end{abstract}

%\begin{keywords}
%cosmology: CMB -- data analysis
%\end{keywords}

\section{Introduction}

It is commonly assumed in cosmology that the Universe is infinite.
While this assumption greatly simplifies calculations, there is nothing forbidding a geometry of space-time with a finite topology. General Relativity specifies the local curvature of space-time but the global geometry still remains undefined. 

The CMB contains a wealth of information about our Universe. One expects that observation of CMB anisotropies will yield information on local geometry parameters such as the expansion rate of the Universe, its dark matter content as well as its nature, luminous matter content, the local curvature, etc. It is also expected that this same source might help  us infer the global topology of the universe since topological scales are on the order of the largest observable scales.
The {\it COBE}-DMR data has been used to place constraints on some non-trivial topologies such as flat and limited open topologies. (A summary of such constraints as well as a brief account of  several methods used is given in Section 3).  Meanwhile an improvement in the sensitivity of CMB experiments has stimulated a renewed interest in the search for the space-time topology.

Here we aim to constrain these finite topologies using spherical Mexican Hat
wavelets. In Section 2 we start by giving a a brief account of the topological
models investigated in our analysis. In Section 3  we give a summary of other
methods used to search for the topology of the universe. In Section 4 we
describe the wavelet-based technique applied in the analysis presented here,
and in particular, how we calculate the correlation between the wavelets maps
and the subsequent statistical analysis. We show that these wavelets allow one
to discriminate between some topologies while simultaneously constraining the
best fit topological size for each of the models, within the range of
topological sizes and scales studied in this work. In Section 5. we present the
methodology used as well as the results. Finally the conclusions are given in
Section 6.

\section{Topology}

Topology is the mathematical framework concerned with studying continuity.
Therefore a topologist does not distinguish between a circle, a square and a
triangle, or between a doughnut and a coffee cup for example. However topology
distinguishes a coffee cup from a bowl. Here we are interested in the study of
the imprints left on the CMB by multiconnected universes. A simply connected
manifold is a surface where any loop (closed path) can be continuously deformed
into a point;  if not the manifold is said to be multiconnected.

Recent high resolution experiments, such as Boomerang, MAXIMA and DASI
(Netterfield et al.\ 2001, Lee et al.\ 2001, Pryke et al.\ 2001), seem to
favour an approximately flat observable universe. Consequently we restrict here
our attention to non-trivial topologies with a flat geometry, and even if
$\Lambda$ models are favoured by those data we will not consider them in the
analysis presented here. For a comprehensive review on topology see among
others Lachieze-Rey \& Luminet 1995, Luminet \& Roukema  1999, Levin 2001. The
description given here follows closely that of Levin, Scannapieco \& Silk 1998
and Scannapieco, Levin, \& Silk 1999.

There are 6 compact orientable flat topologies. As described below, these
topologies can be constructed using a parallelepiped or a hexagonal prism as
the finite fundamental domain whose opposite faces are glued together (ie.
identified) in some particular way. These identifications can be represented by
tiling the space with copies of the fundamental domain. Four of these
topologies are obtained from the parallelepiped and two from the hexagon.

\begin{itemize}

\item Model 1:
The simplest one is the hypertorus (3-torus) which is obtained from a parallelepiped with pairs of opposite faces glued together. In other words this manifold is built out of a parallelepiped by identifying $x \rightarrow x+h$, $y \rightarrow y+b$ and $z \rightarrow z+c$.

The other three manifolds are variations of the hypertorus involving identifications on opposite faces of a twisted parallelepiped. 

\item Model 2: One of these has opposite faces identified with one pair rotated through the angle $\pi$ (here called $\pi$ twist torus); 

\item Model 3: the second identifies opposite faces with one rotated by $\pi/2$ (the $\pi/2$ twisted torus),

\item Model 4 : the third one is obtained by proceeding with the following identifications: $(x,y,z) \rightarrow (x+h,-y,-z)$ corresponding to translation along x and rotation around x by $\pi$; next $(x,y,z) \rightarrow (-x,y+b,-(z+c)) $ corresponding to translation along y and z followed by rotation around y by $\pi$; and finally $(x,y,z) \rightarrow (-(x+h),-(y+b),z+c)$ translation along x,y,z followed by rotation around z by $\pi$ (here called the triple twist torus).

Two other topologies are built out of a hexagon by identifying the three pairs of opposite sides while in the z direction, 

\item Model 5: the faces are rotated relatively to each other by $2\pi/3$ ($2 \pi/3$ hexagon) 

\item Model 6: and by $\pi/3$ (here called $\pi/3$ hexagon).

\end{itemize}

For large scales, the temperature fluctuations in the CMB are mainly due to the potential fluctuations on the last scattering surface, the so called Sachs-Wolfe effect, SW (Sachs \& Wolfe 1967). This effect corresponds to a redshifting of the photon as it climbs out of 
the potential well on the surface of last scattering,
and  also to a time dilation effect which allows us to see them at a different 
time from the  unperturbed photons (Coles \& Lucchin 1995, Efstathiou 1990).
Its expression is given by:
\be
\left(\frac{\Delta T}{T}(\hat{n})\right)_{sw}={\textstyle{\frac{1}{3}} \Phi(\Delta \eta \hat{n})}
\ee
where $\Phi$ is the gravitational potential and $\Delta \eta$ is the conformal time between the decoupling time and today, ie, the radius of the Last Scattering Surface (LSS) in comoving units. The observable universe is defined by the diameter of the LSS $2 \Delta \eta$.
The angular power spectrum is defined as $C_{l}=\langle |a_{lm}|^{2} \rangle $, where the multipole moments $a_{lm}$ are the coefficients of the standard expression of temperature fluctuations of the CMB on the celestial sphere, in terms of an expansion in spherical harmonics:
\be
\frac{\Delta T}{T}(\alpha,\phi)=\sum_{l=2}^{\infty} \sum_{m=-l}^{l} a_{lm} Y_{lm}(\alpha,\phi) \label{eq:dt}
\ee
where $(\alpha,\phi)$ are the polar coordinates of a point on the spherical surface. 
For flat ($\Omega=1$) scale invariant ($n=1$) models with no contribution 
from a cosmological constant the SW effect gives rise to a flat power spectrum, 
i.e., $l(l+1)C_{l}=$constant, where the quantity $l(l+1)C_{l}$ is the 
power of the fluctuations per logarithmic interval in $l$.
Its shape is altered if one assumes a tilted initial power spectrum with a power law, $P(k)=Ak^{n}$ or if one incorporates a cosmological constant, or assumes other than flat models. In the last two cases another source of anisotropy appears, the so-called integrated Sachs-Wolfe effect, ISW, and is due to the fact that potential fluctuations are no longer time independent.

Now the gravitational potential can be decomposed into eigenmodes:
 
\be
\Phi = \int_{-\infty}^{+\infty} d^{3} \vec{k} \hat{\Phi}_{\vec{k}} exp(i \Delta \eta \vec{k} . \hat{n}) 
\ee

In general, inflationary scenarios predict a Gaussian distribution of fluctuations independent of scale. This implies that the $\hat{\Phi}_{\vec{k}}$ are obtained from a Gaussian distribution with $\langle \hat{\Phi}^{*}_{\vec{k}} \hat{\Phi}_{\vec{k^{'}}} \rangle = \frac{2\pi}{k^{3}} P_{\phi}(k) \delta^{3}(\vec{k}-\vec{k'})$, where the brackets indicate an ensemble average and $P_{\phi}$ is the predicted power spectrum. A Harrison-Zeldovich (ie scale invariant) spectrum corresponds to $P_{\phi}=constant$.
For a compact manifold, the identifications on the fundamental domain are expressed in terms of boundary conditions on $\Phi$ (see appendix A). Hence the continuous $\vec{k}$ becomes a discretized spectrum of eigenvalues. 
In general the temperature fluctuations in a compact, flat manifold become:

\be
\frac{\delta T}{T} \propto \sum_{-\infty < k_{x},k_{y},k_{z} < \infty} \hat{\Phi}_{k_{x}k_{y}k_{z}} exp(i \Delta \eta \vec{k}.\hat{n})
\ee
with additional relations on the $ \hat{\Phi}_{\vec{k}}$.
Once these eigenmodes are known one can obtain the angular power spectrum $C_{l}$ (see Levin, Scannapieco \& Silk 1998, for example).
The set of restricted discrete eigenvalues and relations between the coefficients $\Phi_{k_{x}k_{y}k_{z}}$ set by their specific identifications on the fundamental domain, are extensively given in Levin, Scannapieco \& Silk 1998 and Scannapieco, Levin, \& Silk 1999.
For the sake of completeness we give in Appendix A, a list of these solutions.

The topological identifications performed on the fundamental domain alter the CMB power spectrum in that: 1) the boundary conditions give rise to a discretised set of eigenmodes for the power spectrum 2) the compact spaces are anisotropic and all except for the hypertorus are inhomogeneous, giving rise to anisotropic Gaussian CMB fluctuations, hence non-Gaussian fluctuations (Inoue 2001, Ferreira \&  Magueijo 1997) the existence of a cutoff in perturbations at scales larger then the topological scale in a given direction.
Point 2) implies that the correlation function between any two points on the LSS is no longer simply a function of the angular separation.
These finite universes will also exhibit spatial correlations due to geometrical patterns formed by repetition of topologically lensed cold and hot spots. In our analysis we investigate these spatial correlations. 

The simulations presented here are based on the relations listed in Appendix A, and were produced only for the SW effect, that is, taking into consideration the sources of CMB anisotropies on large angular scales alone.
To find out whether the topology scale is an observable, in other words, whether its effects can be seen on the CMB, one compares the topological scale (in-radius) with $\Delta \eta$. Therefore the number of clones of the fundamental domain (the parallelepiped for example) that can be observed in a compact universe may be calculated by the number of copies that can fit within the LSS.
In Fig.1 we show simulated maps for the 6 topologies for a topological scale equal to half of the horizon size, $j=0.5$ ($j\propto k_x$, see appendix A for details) In Fig.2 we plot a Torus with topological scale of $0.1$, equal to and 10 times the horizon size ($j=0.1,1,10$ respectively) as well as an infinite universe for comparative purposes.

%\newpage
\begin{figure}
\vspace{3in}
%\vbox{
%\hbox{
%\psfig{file=map1.5_smoothed.ps,width=1.7in,height=1in,angle=90}
%\psfig{file=map2.5_smoothed.ps,width=1.7in,height=1in,angle=90}}
%\hbox{
%\psfig{file=map3.5_smoothed.ps,width=1.7in,height=1in,angle=90}
%\psfig{file=map4.5_smoothed.ps,width=1.7in,height=1in,angle=90}}
%\hbox{
%\psfig{file=map5.5_smoothed.ps,width=1.7in,height=1in,angle=90}
%\psfig{file=map6.5_smoothed.ps,width=1.7in,height=1in,angle=90}}
%}
\caption{From left to right in the first row: Simulations for a Torus, $\pi$ twist Torus; in the second row: $\pi/2$ twist Torus, triple twist Torus; and in the third row: $\pi/3$ Hexagon, $2\pi/3$ Hexagon, for topological scale equal to half the horizon size (ie j=0.5). The maps are in HEALPix pixelization with Nside=32 and {\it COBE}-DMR resolution.}
\end{figure}

\begin{figure}
\vspace{2.4in}
%\vspace*{-2in}
%\hspace*{-1in}
%\psfig{file=plots1.ps,width=5in,height=7in,angle=0,clip=}
%\psfig{file=plots1.eps,width=3.5in,height=2.4in,angle=0,clip=}
%\vspace*{-2.5in}
\caption{Torus for j=0.1,1,10 ie topological scale of $0.1$, equal, 10 times the horizon size and an infinite universe.}
%\vspace*{-2in}
\end{figure}

\section{Searching for Topology: Other Methods}

The search for topology can generally be divided into two methods,
direct statistical methods and geometric methods. 

Statistical methods can be used to analyze properties ranging from the correlation function to the angular power, $C_l$, spectra of topology simulations and compare them to the known data.
Geometric methods encompass the search for circles in the sky, look for planes or axes of symmetry, search for pattern formation via maps of correlations, for example, a map of antipodal correlations of the CMB maps, to name a few.
The statistical analysis makes use of averages over the sky and therefore might not be sensitive to the inhomogeneity and anisotropy characteristic of these non-trivial topologies.

Ever since data from {\it COBE}-DMR was released, calculations and statistical analyses have been performed to see how consistent a topologically compact universe is with the actual data.  Initial work comparing $C_l$ spectra from topological simulations of compact flat spaces, with {\it COBE}-DMR data seemed to set a lower limit on the topology scale at $80\%$ of the horizon radius ($40\%$ of the horizon diameter) (Stevens, Scott \& Silk 1993, Levin, Scannapieco \& Silk 1998, Scannapieco, Levin \& Silk 1999), still possibly detectable by some of the methods mentioned above, but eliminating the intriguing possibility of a very small topological scale.
A likelihood analysis was performed to compare the Cl's of the model with the full {\it COBE}-DMR range of the angular power spectrum. 
Though compact topologies do not give rise to isotropic temperature fluctuations, therefore any likelihood analysis is bound to be ambiguous.  
While observations of the power spectrum on large angular scales are used to put limits in the minimum scale of the topology, the cosmic variance does not allow us to differentiate between some of these models (for example between a hexagonal prism and a hypertorus). Therefore the angular power spectrum is a poor measure of topology. Several other methods have been put forward, in particular the geometric methods. 

One of the first geometric searches for topology placed the lower limit even higher (see de Oliveira-Costa \& Smoot, 1995). They showed that the most probable hypertorus was 1.2 times larger than the horizon scale. A more anisotropic topology was also considered such as a hypertorus with 1 or 2 dimensions smaller than the present horizon ($T^{1}$ and $T^{2}$ respectively)(de Oliveira-Costa, Smoot \& Starobinsky 1996). Their analysis is based on the fact that an anisotropic hypertorus will exhibit a symmetry plane or a symmetry axis in the pattern of CMB fluctuations.
They developed a smallness statistic 
to measure the level of symmetry that would result in these cases. 
Their analysis constrained $T^{1}$ and $T^{2}$ models to be larger than half the radius of the LSS, in their small dimensions. 
A circle method has been applied afterwards giving pessimistic results. 
   
While for compact flat models the eigenmodes decomposition has been done successfully, its analytical calculation for compact hyperbolic models is actually impossible. Methods to constrain these spaces, include brute force numerical calculation of the eigenmodes (Inoue 1999, Inoue, Tomita \& Sugyiama 2000, Cornish \& Spergel 2000, Aurich 1999, Aurich \& Marklof 1996) the method of images construction of CMB maps (Bond, Pogosian \& Souradeep 1998, 2000) and of course geometric methods.
Using the eigenmodes computed numerically a comparison of the Cl's with {\it COBE}-DMR data alone seems to not rule out any of the CH models (Aurich 1999, Aurich \& Steiner 2001, Inoue 1999, Cornish \& Spergel 2000, Bond, Pogosyan \& Souradeep 2000)
Although a comparison of the full correlation function does not survive consistency according to the method of images analysis,
 the correlation function can be calculated using the method of images
without the explicit knowledge of eigenmodes and eigenvalues (see Bond, Pogosyan \& Souradeep, 2000). 
They find anisotropic patterns and a long wavelength cutoff in the power spectrum. The anisotropic patterns correspond to spikes of positive correlation between a point on the LSS and its image. 
The suppression of power for compact hyperbolic models is in part compensated by the ISW effect and therefore is not so noticeable as in flat models. They used a Bayesian analysis to compare the CH models with the {\it COBE}-DMR data and found that CH models are inconsistent with the data for most orientations. Only a small set of orientations fit
the data  better than the standard infinite models. 
Their analysis showed the inadequacy of using analysis based on the $C_{l}'s$ alone, by pointing out the large error bars on the $C_{l}'s$ due to the large cosmic variance in the models topologically connected, although there is some controversy over the interpretation of these error bars (Inoue 2001), and the finding of a spatially dependent variance due to the global breaking of homogeneity.
Some other work explored the idea of a correct orientation for the CH model and used the cold and hot spots of {\it COBE}-DMR to infer the orientation of this manifold (Fagundes 1996, 2000).

As pointed out these direct methods are model dependent in that they depend for example of the choice of manifold, the location of the observer, the orientation of the manifold, the local parameter values and finally of the statistics used.  
 
Another of the geometric methods for searching for a finite topology is to look for circles in the sky (Cornish, Spergel \& Starkman 1998).  
These circles would be a matched pair, not circles of identical temperature in the sky, but circles of identical temperature fluctuations.  Each topology would have its own distinct pattern of matched circles, and presumably if all the circle pairs could be identified, the topology of the universe could be inferred from there. This method applies to all multiconnected topologies and it is model independent. To get the correlated circles a statistical approach is used. Work was then done to include the Doppler effects in the analysis (Roukema 2000). Using {\it COBE}-DMR data this method has also been applied to constrain asymmetric flat 3-spaces (Roukema 2000).  
Intriguing as this possibility is, it has its difficulties.  
  Circles are obscured in practice however  by the ISW effect, which since it occurs after the last scattering surface would not be correlated in the circles, and by the fact that the galactic cut will interfere with many of the circles, making them more difficult to detect (Levin 2001).  Even if these obstacles were surmounted, this method is useless if the topological scale is larger than the horizon size, for if the surfaces of last scattering of the observer do not intersect, there are no circles.  
Cornish and Spergel applied this test to simulated maps to conclude that the ISW interfered in the statistic giving rise to a poor match for circle pairs. This effect affects mainly the low order multipoles and therefore an improvement is to be expected from future satellite missions.
Bond Pogosian and Souradeep found good matches even at {\it COBE}-DMR resolution, but found that at low values of the density parameter $\Omega_{0}$ when the ISW effect contribution is larger the correlations along circles get worse.
Hence the {\it COBE}-DMR is not useful
for detecting  circles in  hyperbolic universes since the ISW will contribute to the multipole range covered by {\it COBE}-DMR.
However this approach  might still be useful for flat spaces without a cosmological constant.
   
Another geometric method that has been used to search for possible topology is to look for pattern formation.  Periodic boundary conditions imposed by the topological identifications would lead to the formation of distinct patterns at individual modes on the last scattering surface.  These patterns are not readily apparent because the superposition of many modes drowns out what would be obvious in one mode. Looking for correlations is one of the best methods to search for patterns. 
An antipodal correlation, as described in Levin et al. 1998,
ie, correlation between points on opposite sides of the last scattering surface,
would give a monopole, that is, no correlation at all, in a simply connected universe since these points should be out of causal contact.  In a multi-connected universe though, they could in fact be very close to each other (even may be the same point) and show significant correlation. Hence this technique searchs for ghost images and in doing so uncovers the symmetries of space. These correlated maps satisfy a desirable property since they are model independent.  
In theory it would then be possible to extrapolate from the patterns and symmetries in the anti-podal maps to the global topology.  
The resolution of the spots needed for the correlation calculation is far higher than that obtained by {\it COBE}-DMR, somewhere on the Silk damping scale, but this could be obtained by MAP or Planck (Levin \& Heard, 1999).  This method runs into the same problems as the circles in the sky method, particularly with regards to noise and disentangling correlations on the last-scattering surface from the uncorrelated ISW and foreground effects that obscure it. A real space statistic is needed to overcome these obstacles,
as proposed below.

\section{Wavelets as scale information providers}

We propose to apply a different method which makes use of wavelets to detect these patterns on the CMB sky. It has been shown that this tool is quite helpful in assessing the contamination of CMB data by discrete radio sources, as well as in detecting and characterizing non-Gaussian structure in maps of the CMB. These functions are widely used in analysis and compression of data because they are computationally efficient and have a better localization in both space and frequency than the usual Fourier methods.

Wavelet deconvolution of an image provides information at each location
of the contribution of different scales. It is our aim in this paper to 
distinguish topologies through the imprint left on the 
CMB. The geometry of space will introduce spatial correlations in the CMB and thereby
 generate patterns in the observed maps. 
Full use of all the information provided by wavelets would consist of 
interpreting the localized values of wavelet coefficients at each scale.
Our approach will however be a statistical one. The information gathered 
at each scale and each location will be combined into the scale-scale 
correlation between scales $k$ and $k'$ defined as 
$$C(k,k')=(1/N_p)\sum_{i=1}^{N_p}[\Delta T/T(k,i)][\Delta T/T(k',i)])$$ 
\noindent where $N_p$ represents the number of pixels. $\Delta T(k,i)$ is
the value of the wavelet coefficient at pixel $i$ after convolution with
a wavelet of size $k$. That is, if pixel $i$ corresponds to direction $\vec b$,
then
$$\Delta T/T(k,\vec b)=\int d\vec x \Delta T/T(\vec x) \Psi(k,\vec b;\vec x),$$
\noindent where $\Psi(k,\vec b;\vec x) =(1/k)\psi(\vert \vec x-\vec b\vert /k)$ 
and $\psi$ is the so called ``mother'' wavelet.

Planar Daubechies wavelets have been implemented in the HEALPix pixelization\footnote{http://www.eso.org/science/healpix/}(G\'{o}rski, Hivon \& Wandelt 1999), and applied to
projections of simulated CMB sky maps of flat models with a non-trivial
global topology (Canavezes et al.\ 2000). Their results are consistent with our own, shown in section 5. 
Only two spherical 
wavelets have been up to now implemented on the HEALPix pixelization. 
The spherical Haar wavelet has been used in the quad-cube pixelization 
(Tenorio et al. 1999) as well as in the HEALPix one (Barreiro et al. 2000).
The spherical Mexican Hat was implemented on the HEALPix pixelization by
Cay\'on et al.\ 2001 and Mart\'\i nez-Gonz\'alez et al.\ 2001 (details about the spherical Mexican Hat wavelet can be found in those papers). 
The scales that can  be studied by the Haar wavelet are powers of two times 
the pixel size. The scale-scale correlation function therefore will not be
calculated in evenly discretized scales. On the contrary, the 
spherical Mexican Hat allows us to determine wavelet coefficients at 
any scale, although successive scales will not be independent. 
Moreover, Mart\'\i nez-Gonz\'alez et al. 2001 have shown 
that this wavelet is better suited to detect certain non-Gaussian features 
than the spherical Haar.

The scale-scale correlation (defined as above or as the correlation of the 
squared values of $\Delta T/T$)
has been previously used to search for non-Gaussianity in the {\it COBE}-DMR data (Barreiro et al. 2000, Cay\'on et al. 2000). {\it COBE}-DMR values of this quantity are
found to be consistent with Gaussianity. 
Previous efforts have been dedicated to comparing 
the power spectrum of the {\it COBE}-DMR data with predictions from 
different topologies as mentioned in the previous section. 
The power
spectrum can be viewed as the correlation of two equal scales.  
Power spectra for
flat topologies such those studied in this paper are presented 
in Scannapieco et al. 1999. 

%\newpage
\begin{figure}
\vspace{2in}
%\hspace*{-0.5in}
%\hbox{
%\psfig{file=mapsmall_smooth.eps,width=1.7in,height=1in,angle=90}
%\psfig{file=map1b_290.eps,width=1.7in,height=1in,angle=90}}
%\hbox{
%\psfig{file=map1b_470.eps,width=1.7in,height=1in,angle=90}
%\psfig{file=map1b_830.eps,width=1.7in,height=1in,angle=90}}
\caption{Simulation of a Torus with j=0.1 ie with a topological scale $10\%$ of the horizon size (left hand side), all the others correspond to the simulated map convolved with the Mexican hat wavelet for scales (from left to right): 290,470,830,arcminutes.}
\end{figure}

\begin{figure}
\vspace{2in}
%\hbox{
%\psfig{file=map1_smooth.eps,width=1.7in,height=1in,angle=90}
%\psfig{file=map10b_290.eps,width=1.7in,height=1in,angle=90}}
%\hbox{
%\psfig{file=map10b_470.eps,width=1.7in,height=1in,angle=90}
%\psfig{file=map10b_830.eps,width=1.7in,height=1in,angle=90}}
%%\caption{}
\caption{Simulation of a Torus with j=1 ie with a topological scale of the order of the horizon size (left hand side), all the others correspond to the simulated map convolved with the Mexican hat wavelet for scales (from left to right): 290,470,830 arcminutes.}
\end{figure}

\section{Results}
The simulations presented here consider only the SW effect.
The CMB temperature fluctuations are obtained using Equation 4 of Section 2, for the restricted set of discrete eigenvalue spectra. There are additional relations between the coefficients of the eigenmodes (listed in Appendix A), set by the specific identifications on the fundamental domain. 
The simulations consider topologies with equal-sided fundamental domains, $\Omega_{\Lambda}=0$ and a Harrison-Zeldovich Gaussian power spectrum (see Section 2). 
We simulate a whole sky topology map in HEALPix pixelization which is then convolved with the {\it COBE}-DMR beam (Wright et al.\ 1994). 
The {\it COBE}-DMR noise is then added to the simulated map and a galactic mask (as defined in Banday et al. 1997) is applied. After monopole and dipole removal the map is renormalized to the {\it COBE}-DMR value of $C_{10}$. The simulated map is then convolved with the Mexican Hat wavelet coefficients for a set of scales. 
Due to the high computational cost of each topology simulation we present statistical results for only 100 simulations
in each case. The error we introduce is studied in the case of model 2, for $j=1.5$  for which we ran 500 simulations (see comments later in this Section). 
In Fig.3 and Fig.4 we plot the wavelet maps, ie the topology maps convolved with a Mexican Hat wavelet, at different scales. 
We then compute the scale-scale correlations between successive scales of such maps.
In Figs. 5 to 10 we plot these scale-scale correlations of the 6 topologies for 3 of the 5 different topological scales considered j=0.5,0.8,1.0,1.5,2.0 (ie. topological scale 0.5,0.8,1,1.5,2 times the horizon size respectively), for Gaussian simulations of an infinite universe and its values for the {\it COBE}-DMR data. 

For most of the topologies the shape of the curve changes with topological scale, with its peak moving towards larger angular scales with increasing topological scale. 
The peak possibly arises due to a cutoff of power at large scales (corresponding to the low $l$ cutoff in the $C_{l}$). This feature however
allows the wavelet analysis to distinguish characteristic scales for most of the models. 
For small universes these curves differ substantially from the standard infinite model curve, hence allowing us to discriminate between them.
As the universe becomes bigger the shape of the correlation curves tend to that of an infinite Gaussian universe as should be expected. Nevertheless the mean value of the correlations, hence the  cosmic variances, for a large universe does not tend to that of an infinite universe, rather being substantially larger. The cosmic variances are smaller for a small universe as expected.
This might indicate that we did not consider a large enough upper limit on the topologial sizes of the universes considered. It is to be expected that as the size of the universe increases the mean value and hence the error bars of the scale-scale correlations decrease and tends to that of an infinite universe. This might become apparent for sizes larger then twice the radius of the LSS which is the upper limit of the range of sizes considered here.

The triple twist torus seems to behave in a quite distinct way when compared with the other models. The scale-scale correlation peaks towards the highest angular scale for all topological sizes with roughly the same shape but with similar uncertainties. Therefore this analysis does not allow 
one to differentiate topological scales for this model as easily. This behaviour might result from the limited range of scales studied. The wavelet analysis has  only been applied for scales up to 1800 arcminutes. It might happen that the triple twist torus exhibits a correlation curve shape similar to that of the other models peaking towards scales larger then our upper limit. This would explain why we do not see such a peak. For this particular model one might need to go to larger topological sizes to get similar results to the other models, although this seems to be in variance with the analysis based on the power spectrum which indicates that all 6 compact, orientable and flat topologies are cut off at the same wavelength as the hypertorus (Levin, Scannapieco \& Silk 1998). 

To estimate the error introduced in our analysis by the use of 100 simulations, we plot in Fig. 11 the scale-scale correlation curve and its $1\sigma$ error bar for 100 and 500 simulations. This plot shows that a larger number of simulations  slightly increases the mean value of the scale-scale correlation and increases the variance, although these differences are small when compared with the error bar size and therefore for intercomparative purposes a number of 100 simulations is sufficient to draw the relevant conclusions. 

We then compare these  scale-scale correlations for our topology simulations with the values for {\it COBE}-DMR as well as with the scale-scale correlations for a Gaussian model (ie, infinite universe with a trivial topology).
The data is compared with the simulations by performing a $\chi^{2}$ statistic:
\begin{equation}  
\chi^{2}_{k,k'}(i,j))=(C_{i,j}(k,k') - C_{cobe}(k,k') )^{2}/\sigma_{i,j}^{2}(k,k'),
\end{equation}
and $\chi^{2}(i,j)=\sum_{k,k'} \chi^{2}_{k,k'}(i,j)$, 
where $C_{i,j}(k,k')$ is the scale-scale correlation averaged over the 100 simulations for topological model $i$ with topological size $j$, between scales $(k,k')$; $C_{cobe}(k,k')$ is the observed {\it COBE}-DMR value; and $\sigma_{i,j}(k,k')$ is the standard deviation for model $i$ with topological size $j$, for scales $(k,k')$ 
computed from the 100 simulated CMB maps. 

The presence of foregrounds in the {\it COBE}-DMR maps might substantially affect the calculation of these correlations.
To tackle this issue we compute the scale-scale correlation for the {\it COBE}-DMR coadded map after foreground (ie. {\it COBE}-DIRBE) correction. These correlations for the {\it COBE}-DMR both before and after foreground correction are plotted in Fig.12, along with the correlations for a torus (model 1) and an infinite universe. From Fig.12 we conclude that the differences between the two {\it COBE}-DMR curves are too small to be noticeable in our statistical analysis. We do not expect the foregrounds correction to affect our final results.

In order to quantify these results, we compare the mean scale-scale correlation and the $1 \sigma$ scatter with the values of the scale-scale correlation for each simulation via the $\chi^{2}$ values. We get the distribution of the $\chi^{2}$ for each model and each topological size.
We then compute the probability of getting a $\chi^{2}$ value larger than
or equal to that of the {\it COBE}-DMR value, 
and the results are displayed in Table 1.
For all but the triple twist torus, the smallest universe with $j=0.5$ is excluded. For the triple twist torus (model 4), however, this is the favoured case with a probability of $70\%$.
The hypertorus (model 1) with a universe size $80\%$ of the horizon size has a probability of $47\%$ while the $\pi$ and the $\pi /2$ twist torus (model 2 and 3 respectively) with a size equal to the horizon size has a probability of $53\%$ and $50\%$ respectively. Finally, the $2 \pi /3$ and the $\pi /3$ twist hexagon (models 5 and 6 respectively) with size equal to the horizon size have probabilities of $36\%$ and $38\%$ respectively.

In making use of the $\chi^{2}$ statistic, we assume that the error bars follow a Gaussian distribution. 
To assess the effect of  non-Gaussianity
on these simulations  of the distribution of the scale-scale correlations, we plot in Fig.13 some histograms for a scale of 500 arcminutes for models $\pi$ and triple twist torus, to conclude that indeed these are not Gaussian-distributed.
This might slightly bias the scale-scale correlation to larger values.
Hence assuming symmetric error bars results in a slightly worse fit for the larger universes considered here.

The best fit topological scale varies with the model but in the majority a value of $j=1.0$ is favoured, ie a compact flat universe with a topological scale approximately equal to the horizon size.
Although the data seems to prefer a triple twist torus with a topological size of $j=0.5$, i.e. a universe half the horizon size, it is known that the wavelet analysis is not a good scale discriminator for this model. This result probably just reflects the fact that in the $\chi^{2}$ analysis comparable $1\sigma$ error bars for all sizes are used while the average value for a universe with half the horizon size is closer to the {\it COBE}-DMR values. These error bars are comparable to the larger uncertainties for the other models and this would explain the smaller $\chi^{2}$ value obtained.
Apart from this model, the data seems to prefer a topological size equal to the horizon size for all but the hypertorus model. For the hypertorus, a topological scale of $j=0.8$ ie with a universe size $80\%$ of the horizon size is preferred, with a $47 \%$ probability. 
Therefore our analysis seems to be in agreement with previous power spectrum analysis (Stevens, Scott \& Silk 1993, Scannapieco, Levin \& Silk 1999), although our favoured topological sizes are curiously close to the minimum value allowed by the power spectrum analysis alone.

The most intriguing result is that for a triple twist torus: according to the $C_{l}$ analysis, a  model as small as half the radius of the LSS is excluded while here it is favoured. As mentioned above, this is mainly due to the large cosmic variance and the similarity of the shape of the correlation curve for all topological sizes. Actually the infinite universe correlation curve is compatible with this model and the $\chi^{2}$ analysis favours the finite universe mainly due to a larger cosmic variance.
With the finite universes displaying a larger cosmic variance, it is not surprising that we get a better $\chi^{2}$ fit for these models.
Therefore the cosmic variance presents a serious problem when we try to compare these models with an infinite universe, but still allows us to distinguish some of these non-trivial models.
These results were also obtained for a limited range of topological sizes. 

Another point of interest is to investigate how a mismatch of the assumed power spectrum compared to the actual one can lead to incorrect conclusions about the preferred topology. One should expect that changing the amplitude and shape of the power spectrum would change the amplitude and shape of the scale-scale correlations, hence resulting in a different $\chi^{2}$ fit. 

Cay\'on et al. found that the {\it COBE}-DMR data was consistent with Gaussian universes.
Here we find that finite models with the appropriate topological size are as compatible with {\it COBE}-DMR data as an infinite universe (see table 4).

\begin{table}
\begin{center}
\begin{tabular}{|lccr|}
Model	& $j$		& $\chi^{2}$	& Prob($\%$) \\ \hline
1       & 0.5 		& 321.11 	& 0     \\ 
1       & 0.8 		& 7.13 		& 47    \\
1       & 1.0 		& 14.87 	& 20     \\
1       & 1.5 		& 15.16 	& 19     \\
1       & 2.0 		& 26.29 	& 15     \\ \hline
2	& 0.5 		& 1017.09 	& 0	\\
2	& 0.8 		& 41.05 	& 6	\\
2	& 1.0 		& 7.32 		& 53	\\
2	& 1.5 		& 17.79 	& 27	\\
2	& 2.0 		& 12.41 	& 14	\\ \hline
3	& 0.5 		& 188.46 	& 2	\\
3 	& 0.8 		& 34.76 	& 12	\\
3 	& 1.0 		& 8.51 		& 50	\\
3 	& 1.5 		& 15.58 	& 27	\\
3 	& 2.0 		& 21.93 	& 16	\\ \hline
4	& 0.5 		& 4.14 		& 70	\\
4	& 0.8 		& 7.62 		& 39	\\
4	& 1.0 		& 14.55 	& 21	\\
4	& 1.5 		& 24.80 	& 16	\\
4	& 2.0 		& 16.56 	& 21	\\ \hline
5	& 0.5 		& 42.68 	& 06	\\
5 	& 0.8 		& 50.28		& 6	\\
5 	& 1.0 		& 14.06 	& 36	\\
5 	& 1.5 		& 16.04 	& 24	\\
5 	& 2.0 		& 21.70 	& 12	\\ \hline
6	& 0.5 		& 116.78 	& 1      \\
6	& 0.8 		& 10.01 	& 38	\\
6	& 1.0 		& 9.60		& 38	\\
6	& 1.5 		& 16.32 	& 21	\\
6	& 2.0 		& 30.97 	& 13	\\ \hline
Gaussian& -		& 10.83		& 45 	\\ \hline
\end{tabular}
\end{center}
\caption{{\it COBE}-DMR $\chi^{2}$ values listed on third column. The last column displays the probability of obtaining such value or higher for each model and each topological size.}
\end{table}

\section{conclusions}

In this paper, we have used spherical Mexican Hat wavelets to study the scale-scale correlations exhibited by non-trivial topologies. Wavelet deconvolution of an image provides information about the contribution of different scales at each location in the map. Here we perform a statistical analysis by combining the information gathered at each location and each scale into the scale-scale correlation between two successive scales as defined in Section 4. We compute the scale-scale correlations for 100 simulations of flat compact topologies. These values were compared to the corresponding quantities for an infinite topology and the values obtained for {\it COBE}-DMR data. To quantify and extract constraints we perform a $\chi^{2}$ analysis of the {\it COBE}-DMR data. 
The $\chi^{2}$ values obtained for the preferred topological sizes when compared with an infinite universe might occur due to the large error bars displayed by the scale-scale correlations of these models and not because these spaces do actually fit better the data. If correct this would mean that we are cosmic variance limited. The wavelets technique does not seem to be a good scale discriminator for a triple twist torus for the range of topological sizes considered here, and one might have to go to larger universe sizes to get useful information on this particular model. 
Apart from this model, the data seems to prefer a topological size equal to the horizon size for all but the hypertorus model. For the hypertorus, a topological scale of $j=0.8$ ie with a universe size $80\%$ of the horizon size is preferred, with a $47 \%$ probability. The $\pi$ and the $\pi /2$ twist torus with a size equal to the horizon size is preferred with a probability of $53\%$ and $50\%$ respectively. Finally the $2 \pi /3$ and the $\pi /3$ twist hexagon with size equal to the horizon size is favoured with a probability of $36\%$ and $38\%$ respectively. 
These results need to be interpreted in view of the topological size and scales range used in this analysis.  

One expects that methods based on  a pattern formation technique will help 
us to further constrain these models, in particular with forthcoming satellite experiments such as MAP and Planck. 
In ongoing work we are including the small angle sources of CMB temperature fluctuations in simulating the non-trivial topologies to allow a similar analysis of Planck or MAP data, among others. 

\section{ackowledgments}
We would like to thank Janna Levin and Evan Scannapieco for fundamental comments on the simulations of the non-trivial topologies and Lance Miller, Proty Wu, Pedro Ferreira, Dmitri Novikov and Pedro Avelino for useful discussions. We would also like to aknowledge the use of the HEALPix pixelization.
GR would like to acknowledge a Leverhulme fellowship at the University of Cambridge.
GR and LC thank the Dept. of Physics of the University of Oxford for support and hospitality during the progression of this work.   

%\newpage
\begin{figure}
\hspace*{0.2in}
\psfig{file=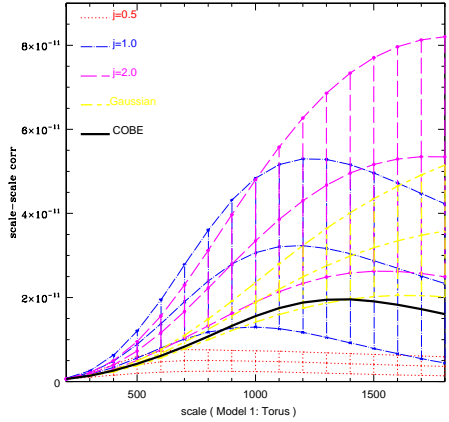,width=2.5in,height=2.5in,angle=0}
\caption{Mean $\pm \ 1 \sigma$ scale-scale correlations for Model 1}
\end{figure}

\begin{figure}
\hspace*{0.2in}
\psfig{file=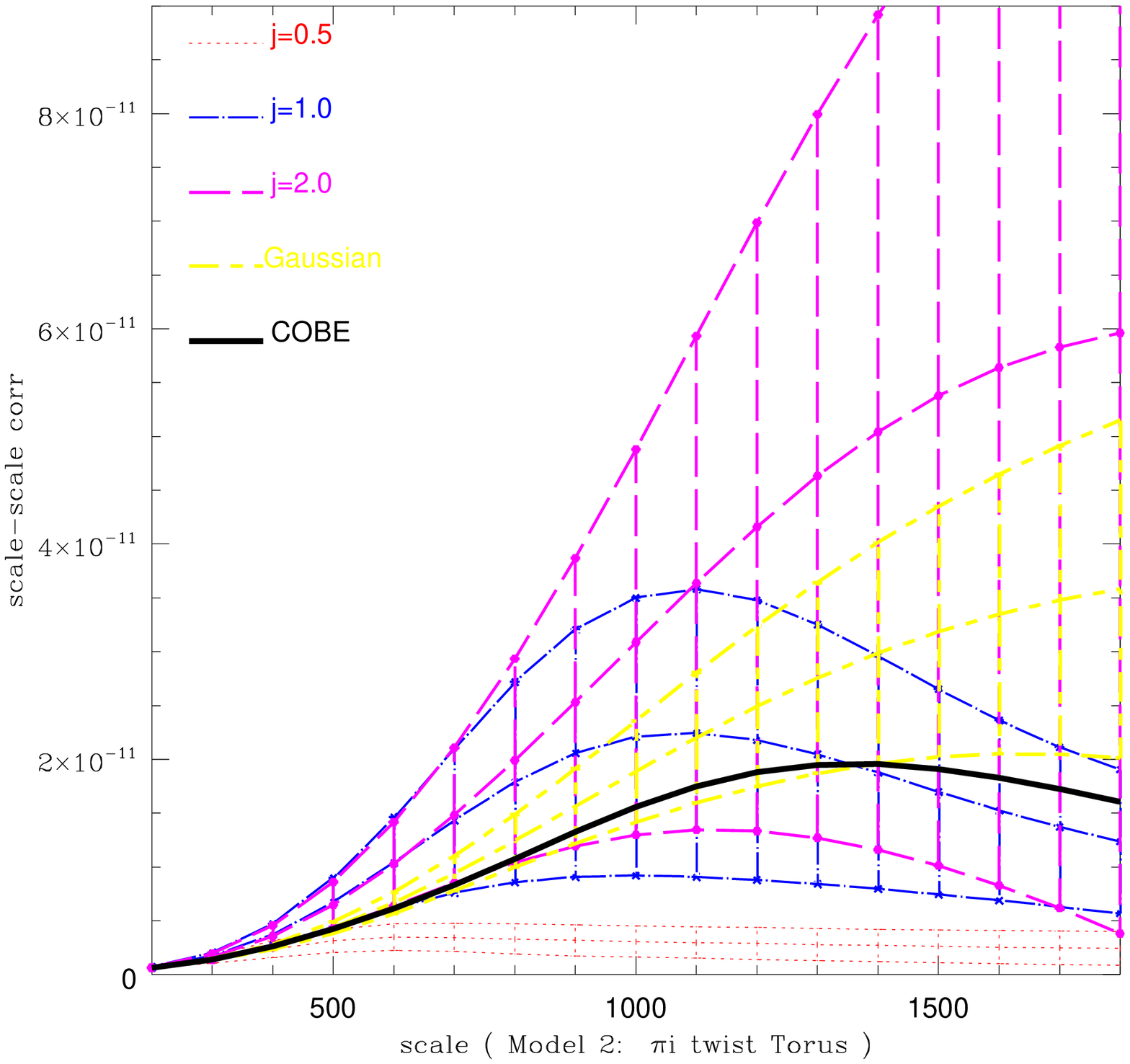,width=2.5in,height=2.5in,angle=0}
\caption{Mean $\pm \ 1 \sigma$ scale-scale correlations for Model 2}
\end{figure}

\begin{figure}
\hspace*{0.2in}
\psfig{file=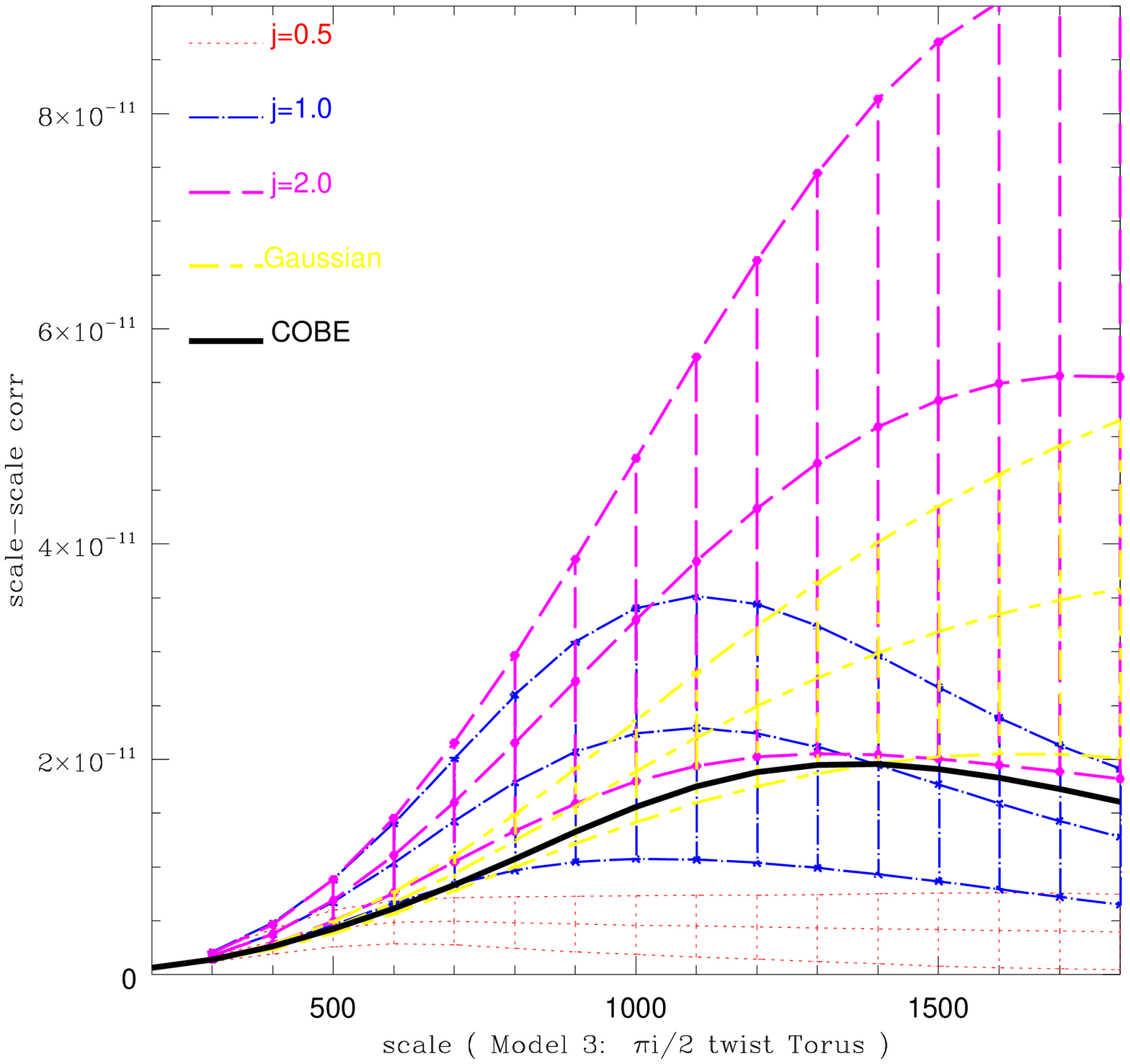,width=2.5in,height=2.5in,angle=0}
\caption{Mean $\pm \ 1 \sigma$ scale-scale correlations for Model 3}
\end{figure}

\begin{figure}
\hspace*{0.2in}
\psfig{file=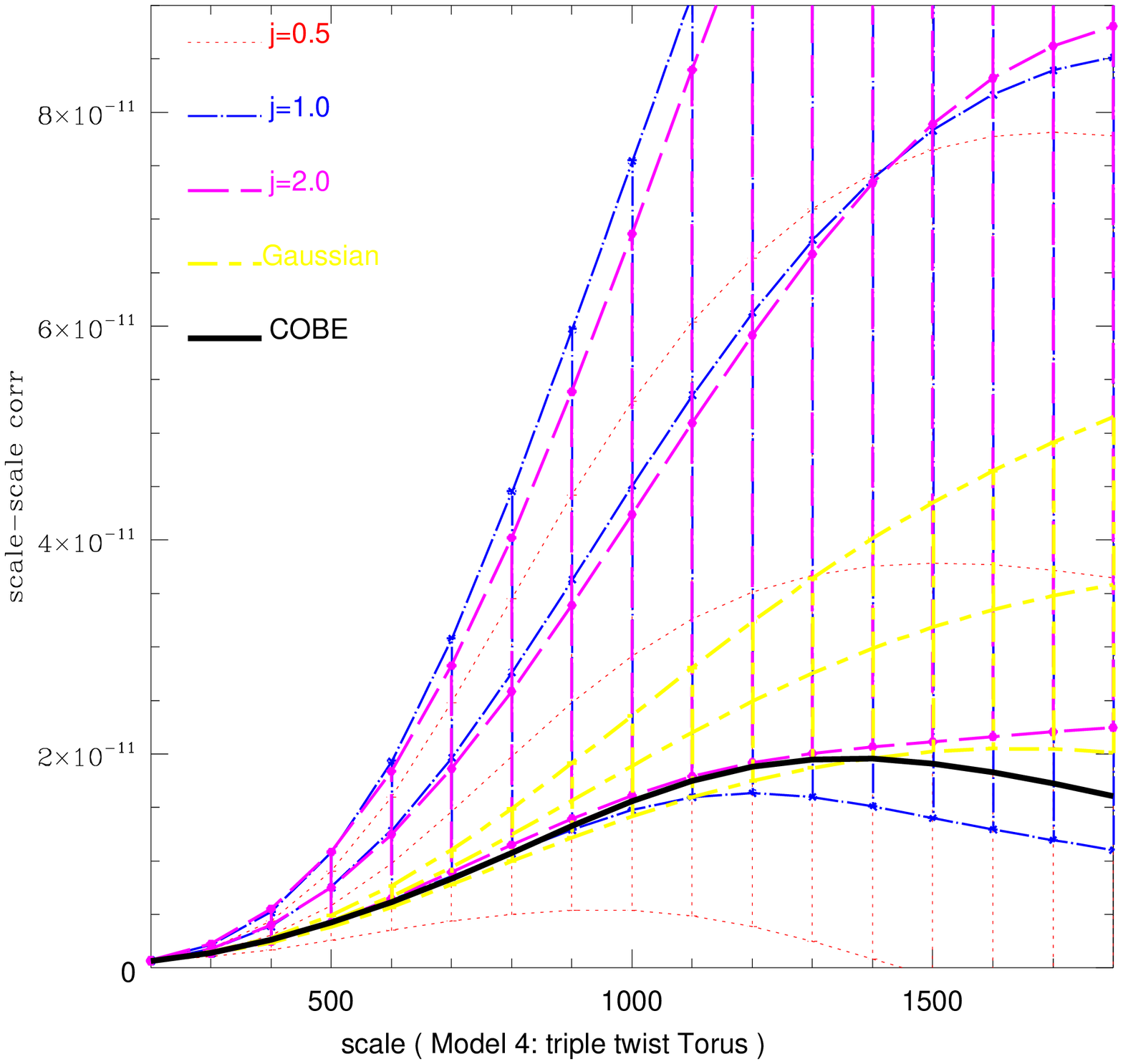,width=2.5in,height=2.5in,angle=0}
\caption{Mean $\pm \ 1 \sigma$ scale-scale correlations for Model 4}
\end{figure}

\begin{figure}
\hspace*{0.2in}
\psfig{file=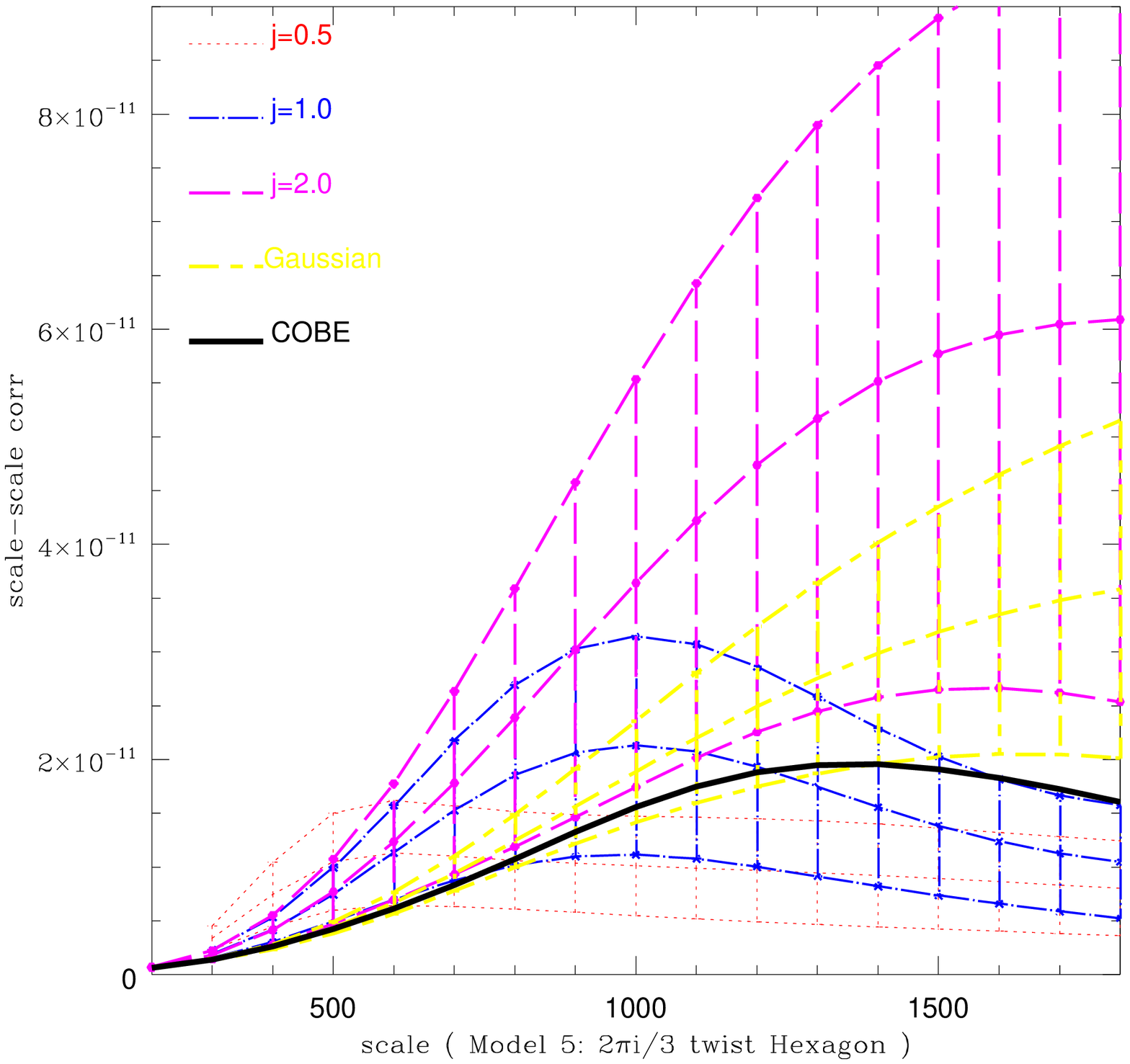,width=2.5in,height=2.5in,angle=0}
\caption{Mean $\pm \ 1 \sigma$ scale-scale correlations for Model 5}
\end{figure}

\begin{figure}
\hspace*{0.2in}
\psfig{file=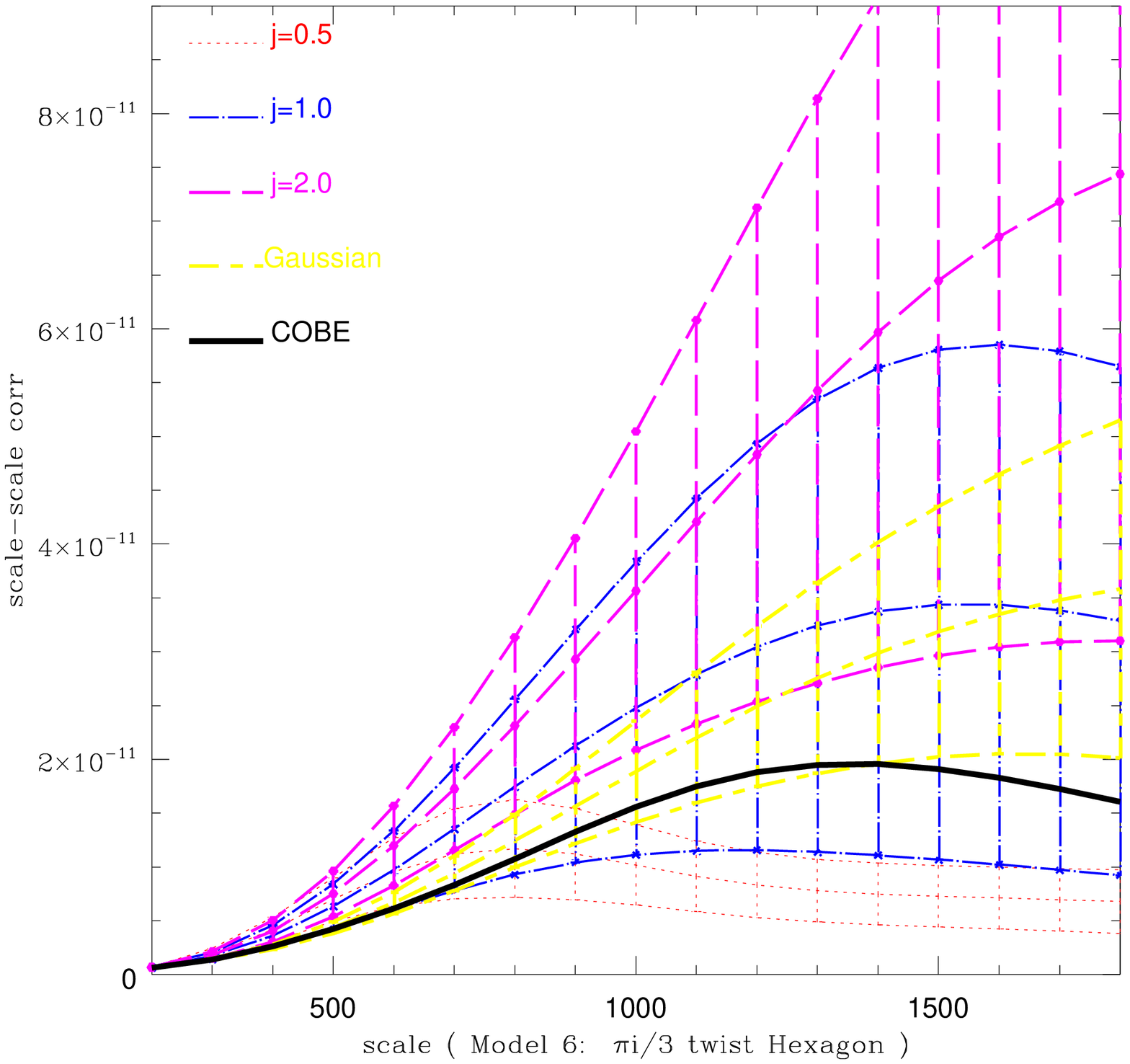,width=2.5in,height=2.5in,angle=0}
\caption{Mean $\pm \ 1 \sigma$ scale-scale correlations for Model 6}
\end{figure}

\begin {figure}
\hspace*{0.2in}
\psfig{file=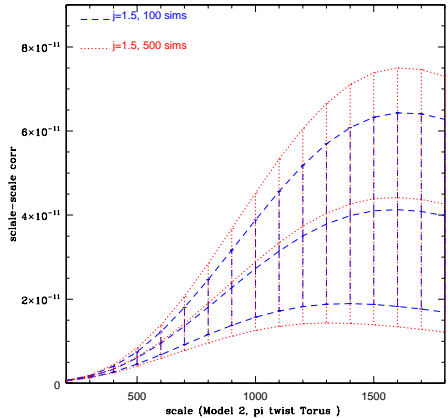,width=2.5in,height=2.5in,angle=0}
\caption{Mean $\pm \ 1 \sigma$ scale-scale correlations for model 2, $j=1.5$ using 100 and 500 simulations.} 
\end{figure}

\begin {figure}
\hspace*{0.2in}
\psfig{file=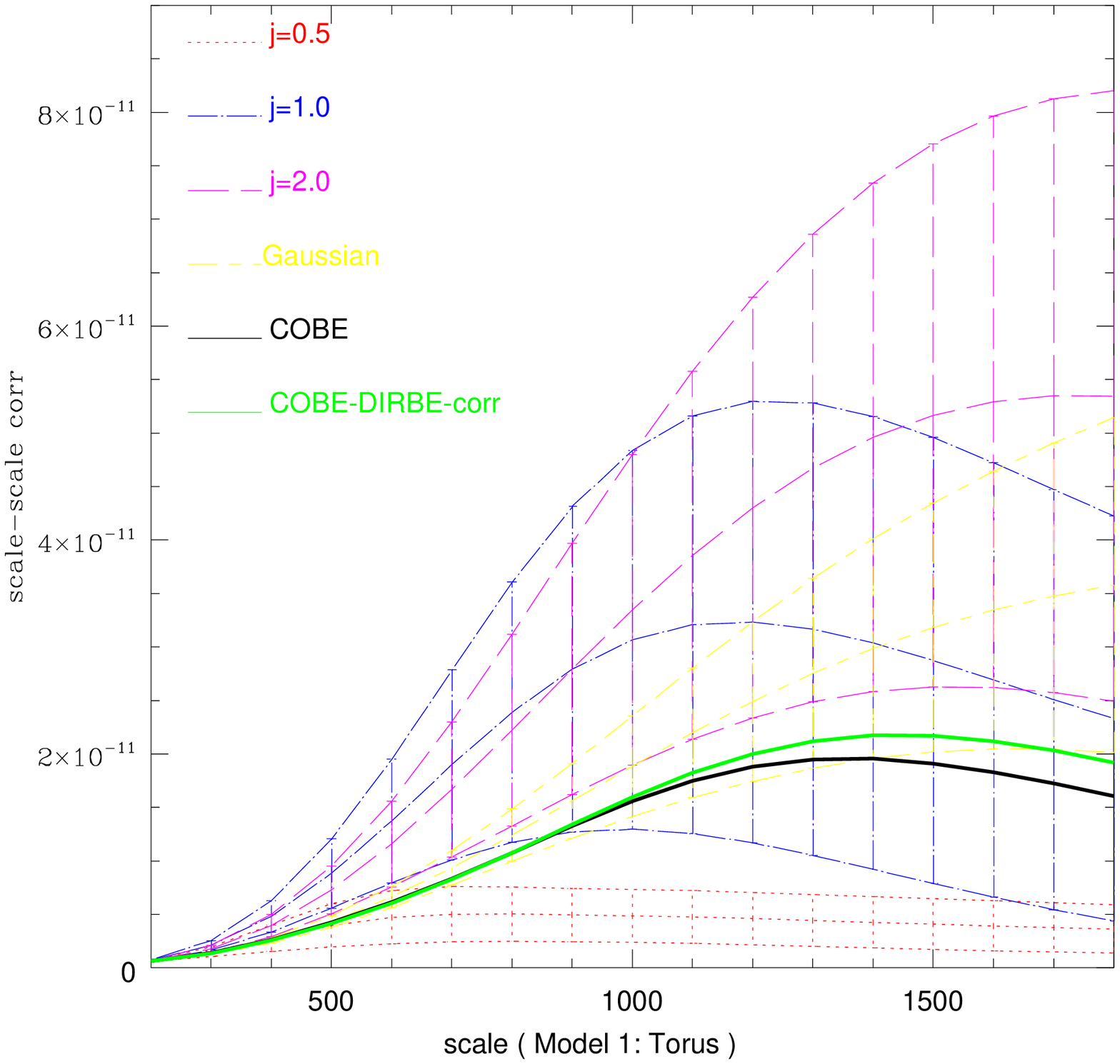,width=2.5in,height=2.5in,angle=0}
\caption{Mean $\pm\ 1 \sigma$ scale-scale correlations for {\it COBE}-DMR before and after foregrounds ({\it COBE}-DIRBE) correction, Model 1 (Torus) and an infinite universe.} 
\end{figure}

\begin{figure}
%\hspace{-0.8in}
\hbox{
\psfig{file=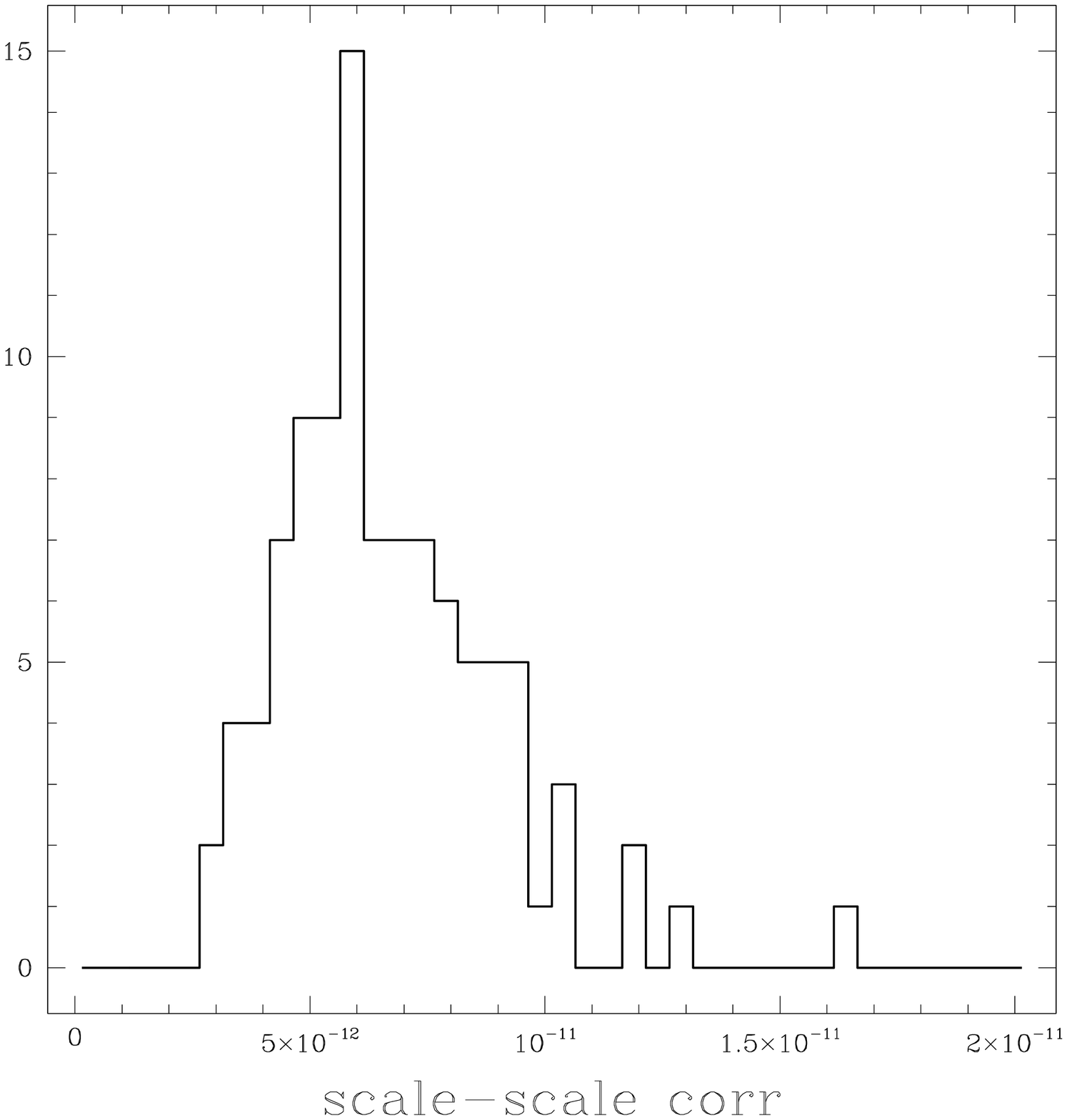,width=1.5in,height=1.5in,angle=0}
\psfig{file=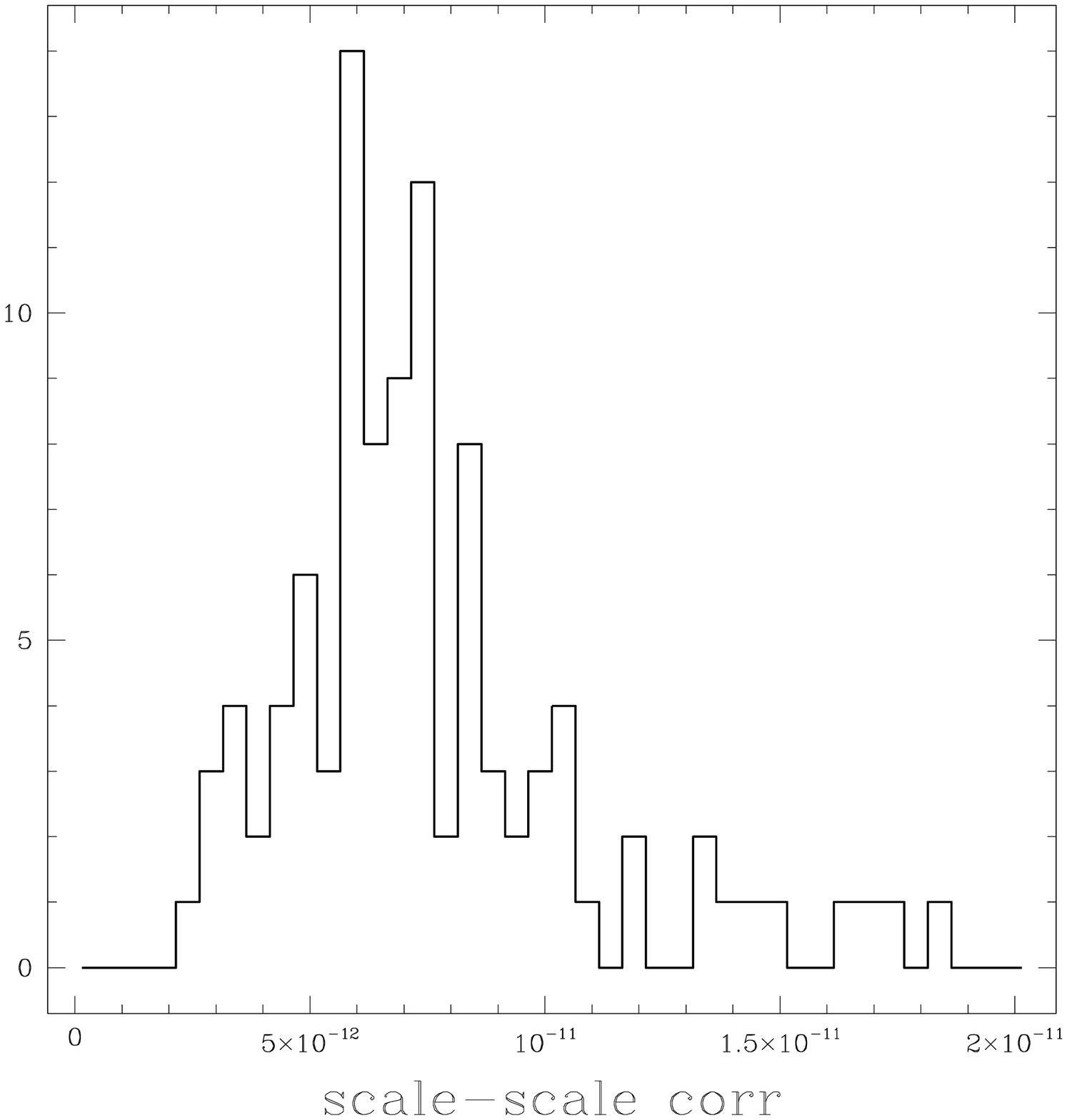,width=1.5in,height=1.5in,angle=0}
}
\caption{Histogram of the scale-scale correlations between 400 and 500 arcminutes for model 2 on the left hand side and model 4 on the right hand side.}
\end{figure}

\section{Appendix A}
In a compact topology the gravitational potential is restricted to a discrete spectra of eigenmodes:
\begin{equation}
\phi(\vec{x})=\sum_{\vec{k}} \phi_{\vec{k}} e^{i\vec{k}.\vec{x}}
\end{equation}
with additional relations imposed on the $\phi_{\vec{k}}$.
For example consider the case of the hypertorus. The identifications on the cube are expressed in terms of three boundary conditions:
\begin{equation}
\phi(x,y,z)=\phi(x+h,y,z)=\phi(x,y+b,z)=\phi(x,y,z+c)
\end{equation}
The first boundary condition gives rise to the following relation:
\begin{equation}
e^{-ik_{x}x}=e^{-ik_{x}(x+h)}
\end{equation}
which imply that $k_{x}=\frac{2\pi}{h}j$. This in conjunction with the two other boundary conditions restrict the set of discrete eigenvalue spectrum to: 
\begin{equation}
\vec{k}=2\pi(\frac{j}{h},\frac{w}{b},\frac{n}{c}), 
\end{equation}
with $j,w,n$ running over all integers. There is a minimum eigenvalue and hence a maximum wavelength fitting inside the fundamental domain given by the paralelepiped: 
$k_{min}=2\pi min(\frac{1}{h},\frac{1}{b},\frac{1}{c})$ and $\lambda_{max}=max(h,b,c)$.
For a $\pi$ twist torus the set of restricted eigenvalue spectrum is
\begin{equation} 
\vec{k}=(2\pi\frac{j}{h},2\pi\frac{w}{b},\pi\frac{n}{c}) 
\end{equation}
with the additional relation on the coefficients of the eigenmodes 
\begin{equation}
\Phi_{jwn}=\Phi_{-j-wn}e^{i \pi n},
\end{equation}
for a $\pi/2$ twist torus these are 
\begin{equation}
\vec{k}=(2\pi\frac{j}{h},2\pi\frac{w}{h},\pi\frac{n}{2c})
\end{equation}
with 
\begin{eqnarray}
\Phi_{jwn}&=& \Phi_{w-jn}e^{in\pi/2}\\
&=&\Phi_{-w-jn}e^{in\pi} \nonumber \\
&=&\Phi_{-wjn}e^{i3n\pi/2}, \nonumber
\end{eqnarray}
while the triple twist torus's discrete spectrum is 
\begin{equation}
\vec{k}=(\pi\frac{j}{h},\pi\frac{w}{b},\pi\frac{n}{c}),
\end{equation}
with 
\begin{eqnarray}
\Phi_{jwn}&=&\Phi_{j-w-n}e^{i \pi j}\\
&=&\Phi_{-jw-n}e^{i \pi (w+n)} \nonumber \\
&=&\Phi_{-j-wn}e^{i \pi (j+w+n)}. \nonumber
\end{eqnarray}
The last two compact flat spaces have an hexagonal fundamental domain.
The potential can be written as:
\begin{eqnarray}
\Phi&=&\sum_{n_{2}n_{3}n_{z}} \Phi_{n_{2}n_{3}n_{z}}e^{ik_{z}z} \\
&\times&  exp[i \frac{2\pi}{h}[n_{2}(-x+\frac{1}{\sqrt{3}}y)+n_{3}(x+\frac{1}{\sqrt{3}}y)]] \nonumber
\end{eqnarray}
For the $2\pi/3$ hexagon one has 
\begin{equation} 
\vec{k}=(2\pi\frac{j}{h},2\pi\frac{w}{h},2\pi\frac{n_{z}}{3c}),
\end{equation}
 with 
\begin{eqnarray}
\Phi_{n_{2},n_{3},n_{z}}&=& 
\Phi_{n_{3},-(n_{2}+n_{3}),n_{z}} e^{i2\pi n_{z}/3} \nonumber \\
&=&\Phi_{-(n_{2}+n_{3}),n_{3},n_{z} e^{i4\pi n_{z}/3}} \nonumber.
\end{eqnarray}
While for the
$\pi/3$ hexagon the spectrum becomes 
\begin{equation}
\vec{k}=(2\pi \frac{j}{h},2\pi \frac{w}{h},\pi \frac{n_{z}}{3c}),
\end{equation}
 with
\begin{eqnarray} 
\Phi_{n_{2},n_{3},n_{z}} &= 
&\Phi_{(n_{2}+n_{3}),-(n_{2}-n_{3})/ \sqrt{3},n_{z} e^{i \pi n_{z}/3}} \\
&=& \Phi_{n_{3},-(n_{2}-n_{3}),n_{z} e^{2i \pi n_{z}/3}} \nonumber \\
&=& \Phi_{-n2,(n_{2}-n_{3}/ \sqrt{3},n_{z} e^{i\pi n_{z}}} \nonumber \\
&=&\Phi_{-(n_{2}+n_{3}),n_{3},n_{z} e^{i 4\pi n_{z}/3}} \nonumber.
\end{eqnarray}


\begin{thebibliography}{}

\bibitem{aurich1} Aurich, R., ApJ 524 497 (1999) 

\bibitem{aurich2} Aurich R., and Marklof J., Physica D92 101 (1996)

\bibitem{aurich3} Aurich R., and Steiner F., MNRAS 323 1016 (2001)

\bibitem{banday}Banday, A.J. et al. 1997, ApJ, 533, 575

\bibitem{belen} Barreiro, R.B., Hobson, M.P., Lasenby, A.N., Banday, A.J., Gorski, K.M. \& Hinshaw, G. 2000, MNRAS, 318, 475

\bibitem{Bond1} Bond, J. R., Pogosyan, D., and Souradeep, T., 2000, Phys.Rev. D62 043005 

\bibitem{bond2} Bond, J. R., Pogosyan, D., and Souradeep, T., 2000, Phys.Rev. D62 042006 

\bibitem{bond3} Bond J.R., Pogosian D., and Souradeep T., in : Proceedings of the XVIIIth Texas Symposium on Relativistic Astrophysics, ed. A. Olinto, J. Frieman, and D.N. Schramm,(World Scientific, 1997) 

\bibitem{bond4} Bond J.R., Pogosyan D., and Souradeep T., in : Proceedings of XXXIIIrd Rencontre de Moriond, 'Fundamental Parameters in Cosmology', Jan. 17-24, 1998, Les Arc, France 

\bibitem{bond5} Bond J.R., Pogosyan D., and Souradeep T., Class. Quant. Grav. 15, 2671, (1998). 

\bibitem{alex}  Canavezes A.,  Rocha G., Silk J., Levin J., [2000], in Proceedings \
of CAPP2000 held at Verbier, Switzerland, 2000.

\bibitem{cayon} Cay\'on, L., Sanz, J.L., Mart\'\i nez-Gonz\'alez, E., Banday, A.J., Arg\"ueso, F., Gallegos, J.E., Gorski, K.M. \& Hinshaw, G. 2001, MNRAS, 326, 1243 

\bibitem{coles95}
	Coles, P., Lucchin, F. 1995, Cosmology: The Origin and
	Evolution of Cosmic Structure, Wiley, Chichester

\bibitem{Cornish1} Cornish, N. J., Spergel, D. N., and Starkman, G. D., 1998, 
                   Class.Quant.Grav. 15 2657-2670 

\bibitem{cornish2} Cornish, N. J., Spergel, D. N., and Starkman, G., Class. Quant. Grav. 15, 2657-2670, (1998) (gr-qc/9602039) 

\bibitem{cornish3} Cornish, N. J., Spergel, D. N., and Starkman, G., Phys. Rev. D 57 (1998) 5982 

\bibitem{cornish4} Cornish, N.J., and Spergel, D.N., Phys. Rev. D 62 (2000), article 087304 

\bibitem{cornish5} Cornish, N.J., and Spergel, D.N., (math.DG/9906017)

\bibitem{cdm2} 
	Efstathiou, G.P., 1990, in Peacock, J.A., Heavens, A.E., 
	Davies, A.T., eds, Physics of the Early Universe, Adam
	Hilger, New York

\bibitem{fagundes1} Fagundes H.V., ApJ 470 43 (1996); Fagundes H.V., astro-ph/0007443

\bibitem{pedro} Pedro G. Ferreira and Jo\~{ao} Magueijo, Phys. Rev. D56 (1997) 4578-4591. (astro-ph/9704052)

\bibitem{healpix} G\'{o}rski, K.M., Hivon, E \& Wandelt, B.D. (astro-ph/9812350) 1999, Proceedings of the MPA/ESO Conference on Evolution of Large-Scale Structure: from Recombination to Garching, 2-7 August 1998; eds A.J. Banday, R.K. Sheth and L. Da Costa, PrintPartners IPSKAMP NL (1999) 

\bibitem{inoue1} Inoue K.T., Class. Quant. Grav. 16, 3071 (1999) 

\bibitem{inoue2} Inoue K.T., Tomita K., snd Sugyiama N., MNRAS 314 L21 (2000)

\bibitem{inoue3} Inoue K.T., Ph.D. Thesis, astro-ph/0103158 

\bibitem{inoue4} Inoue K.T.,Progress of Theoretical Physics, 106 39 (2001) 

\bibitem{inoue5} Inoue, K.T., Class. Quant. Grav. 18, 10, 1967-1977, 2001 (astro-ph/0011462)

\bibitem{lachieze-rey} M.Lachieze-Rey, J.P. Luminet, Phys. Rept. 254 (1995), 135-214. (gr-qc/9605010)

\bibitem{Maxima2} A.T. Lee et al., ApJ. 561 (2001), L1-L6 (astro-ph/0104459) 

\bibitem{Levin1} Levin, J., 2001, Phys. Reports, submitted, gr-qc/0108043; 

\bibitem{Levin2} Janna Levin, Evan Scannapieco and Joseph Silk, Phys. Rev. D58, 1998, 103516 (astro-ph/9802021) 

\bibitem{Levin3} Levin, J., Scannapieco, E., de Gasperis, G., Silk, J., and Barrow, J. D., 1998 (astro-ph/9807206)  

\bibitem{Levin4} Levin, J., and Heard, I., 1999, ``Topological Pattern Formation'', from conference proceedings for the ``Cosmological Topology in Paris'' (CTP98) workshop.

\bibitem{luminet} J.P. Luminet, B.F. Roukema, 1999, 'Topology of the Universe: Theory and Observations', Carg\`{e}se summer school `Theoretical and Observational Cosmology' ed. Lachi\`{e}ze-Rey M., Netherlands: Kluwer,p117. (astro-ph/9901364)

\bibitem{margonz} Mart\'\i nez-Gonz\'alez, E., Gallegos, J.E., Arg\"ueso, F., Cay\'on, L. \& Sanz, J.L. 2001, astro-ph/0111284

\bibitem{Boomerang2} C.B. Netterfield et al., (astro-ph/0104460)


\bibitem{costa1} de Oliveira-Costa, A. and Smoot, G., 1995,  Astrophys.J. 448 477 

\bibitem{costa2} de Oliveira-Costa, A., Smoot, G. and Starobinsky, A., 1996, Astrophys.J. 468 457

\bibitem{dasi} Pryke, C., et al, ApJ. 568 (2002), 46-51 (astro-ph/0104490)

\bibitem{roukema1} Roukema B.F., MNRAS, 312, 712 (2000) 

\bibitem{roukema2} Roukema B.F., Class. Quant. Grav. 17 3951 (2000)

\bibitem{sachswolfe}
	Sachs, R.K., Wolfe, A.M., 1967, ApJ, 147, 73 

\bibitem{scannapieco1} Evan Scannapieco, Janna Levin, and Joseph Silk, MNRAS 303 (1999) 797 (astro-ph/9811226)

\bibitem{stevens} Daniel Stevens, Douglas Scott, and Joseph Silk, Phys. Rev. Lett., 71, 1, 1993

\bibitem{tenorio}Tenorio, L., Jaffe, A.H., Hanany, S. \& Lineweaver, C.H. 1999, MNRAS, 310, 823 

\bibitem{beam} Wright, E.L., Smoot, G.F., Kogut, A., Hinshaw, G., Tenorio, L.,Lineweaver, C., Bennett, C.L., Lubin, P.M., Astrophys. J., 420, 1-8


\end{thebibliography}
\end{document}